\documentclass[11pt,a4paper,twoside,groupcitations]{article}
\usepackage[]{latexsym,amsmath,amssymb}
\usepackage[]{epsfig}
\usepackage{color}
\usepackage{braket}
\usepackage{graphicx}
\usepackage{cite}
\newcommand{\be}{\begin{eqnarray}}
\newcommand{\ee}{\end{eqnarray}}
\def\beq{\begin{equation}}
\def\eeq{\end{equation}}

\newcommand{\expec}[1]{\mbox{$\langle\, #1\,\rangle$}}

\newcommand{\lp}{\ell_{\rm p}}
\newcommand{\mpl}{m_{\rm p}}
\newcommand{\gn}{G_{\rm N}}

\newcommand{\dd}{\mbox{${\rm d}$}}

\title{\bf Minimum length (scale) in Quantum Field Theory,
Generalized Uncertainty Principle and the non-renormalisability of gravity}
\author{Roberto~Casadio$^{ab}$\thanks{E-mail: casadio@bo.infn.it},
$\ $
Wenbin~Feng$^{ab}$\thanks{E-mail: wenbin.feng@studio.unibo.it},
$\ $
Iber\^e Kuntz$^{c}$\thanks{E-mail: kuntz@fisica.ufpr.br}
$\ $
and
Fabio Scardigli$^{de}$\thanks{E-mail: fabio@phys.ntu.edu.tw}
\\
\\
$^a${\em Dipartimento di Fisica e Astronomia, Universit\`a di Bologna}
\\
{\em via Irnerio~46, 40126 Bologna, Italy}
\\
\\
$^b${\em I.N.F.N., Sezione di Bologna, I.S.~FLAG}
\\
{\em viale B.~Pichat~6/2, 40127 Bologna, Italy}
\\
\\
$^c${\em Departamento de F\'isica, Universidade Federal do Paran\'a}
\\
{\em PO Box 19044, Curitiba -- PR, 81531-980, Brazil}
\\
\\
$^d${\em Dipartimento di Matematica, Politecnico di Milano,}
\\
{\em Piazza Leonardo da Vinci 32, 20133 Milano, Italy}
\\
\\
$^e${\em Department of Applied Mathematics, University of Waterloo}
\\
{\em Ontario N2L 3G1, Canada}
}
\date{}
\begin{document}
\maketitle
\begin{abstract}
The notions of minimum geometrical length and minimum length scale are discussed with reference to correlation functions
obtained from in-in and in-out amplitudes in quantum field theory.
Whereas the in-in propagator for metric perturbations does not admit the former, the in-out Feynman propagator
shows the emergence of the latter.
A connection between the Feynman propagator of quantum field theories of gravity and the deformation parameter
$\delta_0$ of the generalised uncertainty principle (GUP) is then exhibited, which allows to determine 
an exact expression for $\delta_0$ in terms of the residues of the causal propagator.
A correspondence between the non-renormalisability of (some) theories (of gravity) and the existence
of a minimum length scale is then conjectured to support the idea that non-renormalisable theories
are self-complete and finite.
The role played by the sign of the deformation parameter is further discussed by considering an implementation
of the GUP on the lattice. 
\end{abstract}
%
%
%
%
%
%
\newpage
\section{Introduction}
\label{S:intro}
\setcounter{equation}{0}
The idea that spacetime is endowed with a minimum (fundamental) length originated
as a possible cure for the ultraviolet (UV) divergences of quantum field theory~\cite{pauli}, and
then regained notoriety with the increasing interest in quantum gravity and trans-Planckian effects
(for a comprehensive review, see Ref.~\cite{Hossenfelder:2012jw}).
Many candidates for quantum gravity exhibit a minimum length, from string theory to
loop quantum gravity.
A minimum length, or scale, can also be shown to arise from the standard Feynman path
integral for time-ordered in-out amplitudes~\cite{Padmanabhan:1985jq},
which are the ingredients for computing $S$-matrix elements from the Lehmann-Symanzik-Zimmermann
formula.
However, these amplitudes are acausal and complex since they are subjected to Feynman boundary conditions.
An observable minimum length in quantum gravity should be real to arbitrary loop orders
and share the statistical properties of an expectation value.
\par
In this respect, it is therefore very important to distinguish between the use of in-out amplitudes and
in-in amplitudes~\cite{Keldysh:1964ud,Jordan:1986ug}, the latter being the objects which admit
a proper statistical interpretation.
These requirements led some of us to study the minimum length using the in-in expectation value
in Ref.~\cite{Casadio:2020hzs}.
The in-in proper distance, which can be directly interpreted as a geometrical length,
was also compared with the in-out proper ``length'',
which cannot be interpreted as a physical distance but sets the length scale of
scattering processes:
the former was shown to vanish quite generally at the coincidence limit, suggesting that
a geometrical minimum length is most likely absent;
the latter evaluated at the coincidence limit acquires a finite
value of the order of the Planck scale under very general assumptions,
indicating that a minimum length scale is very likely to exist.
The implication of these results is that nothing prevents one from going, in principle, through vanishingly
small distances (spacetime is not ``discrete'' like a lattice where points are represented 
by fixed nodes), but scattering experiments cannot reliably distinguish between events taking place
at the Planck (length) scale or below, since any two processes differing only at trans-Planckian scales
would produce the same scattering amplitudes.
\par
A common approach to investigate the consequences of a minimum length (scale) in
quantum mechanics is given by the Generalised Uncertainty Principle(s)
(GUPs)~\cite{Hossenfelder:2012jw,GUP1,GUP2,GUP3,Casadio:2020rsj}.
GUPs, typically derived via {\em gedanken\/} experiments, are usually encoded in modified
commutators for the canonical observables containing free parameter(s) which, in turn,
determine the minimum length.
Since quantum mechanics emerges in the non-relativistic limit of the one-particle sector
of quantum field theory, one can be tempted to draw the origin of the modified quantum
mechanical commutators to modified field commutators.
However, since the emergence of a minimum length scale does not require any modification
of the quantum field dynamics, it appears more natural to assume that the GUP provides
an effective description of scattering processes in suitable regimes.
The expression for the minimum length scale from Ref.~\cite{Casadio:2020hzs} can then be used
to determine the parameter(s) of a GUP for given quantum field theories of gravity.
\par
This paper is organized as follows:
in Section~\ref{S:minlength}, we briefly review the main results from Ref.~\cite{Casadio:2020hzs}
and discuss a conjecture connecting renormalisabilty and (absence of) minimum length;
in Section~\ref{S:gup} we introduce the simplest example of a GUP and identify the minimum
length scale emerging from generic field theories of gravity with the one determined by the GUP;
in Section~\ref{signdelta} we further support our conjecture by discussing the sign
of the deformation parameter when the GUP is formulated on a lattice;
in Section~\ref{conc} we finally draw our conclusions.
\section{Minimum length scale in scattering processes}
\label{S:minlength}
\setcounter{equation}{0}
In the present section, we briefly review the main results of Ref.~\cite{Casadio:2020hzs}, where
we elaborated a model-independent argument for the absence of a minimum geometrical distance,
but the possibility of a minimum length scale.
\par
We first recall that there are different possible boundary conditions in a quantum field theory.
The most prominently one used in particle physics is the Feynman boundary condition,
which reflects the result of a collision process, scattering an initial state $\ket{0_\text{in}}$
to a final state $\ket{0_\text{out}}$.
Transition amplitudes of this type are complex and do not evolve causally.
They are indeed only instrumental to the calculation of cross sections and decay rates,
which are the sought after quantities at colliders.
Retarded boundary conditions, on the other hand, are required when one is interested
in the time evolution of a system (rather than scattering~\footnote{One might nonetheless
notice that all actual measurements are scattering processes (see discussion of the Heisenberg
microscope at the beginning of Section~\ref{S:gup}).}).
In this case, amplitudes are evaluated on the same state $\ket{0_\text{in}}$,
yielding retarded (causal) propagators and real correlation functions.
\par
Geometrical lengths are obviously real quantities.
When the metric is promoted to an operator for the quantization of gravity,
the quantum geometrical length between the points $x^\mu$ and $y^\mu = x^\mu + \dd x^\mu$
must be defined by the in-in amplitude
\begin{equation}
	\ell_\text{in-in}(x,y)
	=
	\sqrt{\braket{0_\text{in} | \dd s^2 | 0_\text{in}}}
	\ .
\end{equation}
Feynman amplitudes cannot be interpreted geometrically or statistically, but they do provide
the scale of the underlying interaction process
\begin{equation}
	\ell_\text{in-out}(x,y)
	=
	\sqrt{\braket{0_\text{out} | \dd s^2 | 0_\text{in}}}
	\ .
\end{equation}
\par
Classically, the line element $\dd s^2 = \bar g_{\mu\nu}\,\dd x^\mu\, \dd x^\nu$, for some fiducial metric $\bar g_{\mu\nu}$,
goes to zero when $\dd x^\mu$ vanishes at coincident points.
However, this limit in a quantum regime is subtler and turns out to depend on the boundary
conditions discussed above.
The reason boils down to the analytical structure of quantum amplitudes, which may develop
singularities such as poles and branch cuts.
In particular, propagators are typically divergent in the coincidence limit $x^\mu \to y^\mu$.
These divergences can sometimes cancel the vanishing classical length
in the numerator, leaving out a finite and non-zero contribution.
Indeed, adopting the exponential parameterization~\cite{Nink:2014yya,Demmel:2015zfa,Nink:2015lmq}
\begin{align}
g_{\mu\nu}
&
=
\bar g_{\mu\rho} \left(e^{\sqrt{\frac{32\,\pi\,\lp}{\mpl}}\,h}\right)^\rho_{\ \nu}
\nonumber
\\
&
=
\bar g_{\mu\nu}
+
\sqrt{\frac{32\,\pi\,\lp}{\mpl}}\,h_{\mu\nu}
+
\frac{16\,\pi\,\lp}{\mpl}\,h_{\mu\rho}\,h^{\rho}_{\ \nu}
+
O\left((\lp/\mpl)^{3/2}\right)
\ ,
\label{eq:param}
\end{align}
where $\lp = \sqrt{\gn\,\hbar}$ and $\mpl=\sqrt{\hbar/\gn}$ are the Planck length and mass,
respectively, we find
\begin{align}
\lim_{x\to y}
\ell_{\text{in-}\tau}^2
&
=
\lim_{x\to y}
\left(
\bra{0_\tau}g_{\mu\nu}\ket{0_\text{in}}\, \dd x^\mu\, \dd x^\nu
\right)
\nonumber
\\
&
=
\frac{16\,\pi\,\lp}{\mpl}\,
\lim_{x\to y}
\left[
\bra{0_\tau}h_{\mu\rho}(x)\,h^{\rho}_{\ \nu}(y)\ket{0_\text{in}}\,
\dd x^\mu\, \dd x^\nu
\right]
\nonumber
\\
&
\equiv
\frac{16\,\pi\,\lp}{\mpl}\,
\lim_{x\to y}
\left[
(G^{\text{in-}\tau})_{\mu\rho\ \ \nu}^{\ \ \ \rho}(x,y)\,\dd x^\mu\, \dd x^\nu
\right]
\ ,
\label{eq:ds}
\end{align}
where $\tau \in \{\text{in,\ out}\}$ and
\begin{align}
	G^\text{ret}_{\mu\nu\rho\sigma} &= (G^\text{in-in})_{\mu\nu\rho\sigma} ,
	\\
	G^\text{F}_{\mu\nu\rho\sigma} &= (G^\text{in-out})_{\mu\nu\rho\sigma} ,
\end{align}
denote the retarded and the Feynman propagators, respectively.
\par
For simplicity, we shall take $\bar g_{\mu\nu} = \eta_{\mu\nu}$.
We stress, however, that our results can be readily generalized to a curved background
by adopting normal coordinates or the Schwinger proper-time representation.
In momentum space, any free propagator can be written as
\begin{equation}
\Delta_{\mu\nu\rho\sigma}(q^2)
=
\sum_i \frac{\hbar\,P^i_{\mu\nu\rho\sigma}}{q^2-m^2_i}
\ ,
\label{eq:momprop}
\end{equation}
where
\begin{equation}
P^i_{\mu\nu\rho\sigma}
=
\alpha_i\, \eta_{\mu\rho}\,\eta_{\nu\sigma}
+ \beta_i \,\eta_{\mu\sigma}\,\eta_{\nu\rho}
+ \gamma_i\, \eta_{\mu\nu}\,\eta_{\rho\sigma}
\end{equation}
is the most general tensorial structure that can be combined into a tensor
of fourth rank and is symmetric in $\{\mu\nu\}$ and $\{\rho\sigma\}$.
The propagator is thus parameterized by $\alpha_i$, $\beta_i$ and $\gamma_i$, whose values
depend on the particular gravitational Lagrangian.
Different integration contours result in different boundary conditions, which in practice are easily
implemented via the $i\,\epsilon$-prescription.
In position space, Eq.~\eqref{eq:momprop} becomes
\begin{align}	
	G^\text{ret}_{\mu\nu\rho\sigma}(x,y)
	&=
	\sum_i \left[-\frac{\theta(x^0-y^0)}{2\,\pi}\,\delta(\ell^2)
	+ \theta(x^0-y^0)\,\theta(\ell^2)\,
	\frac{m_i\, J_1(m_i\,\ell)}{4\,\pi\,\ell}\right]
	\hbar\,P^i_{\mu\nu\rho\sigma}
	\ ,
	\\
	G^\text{F}_{\mu\nu\rho\sigma}(x,y)
	&=
	\sum_i \frac{\hbar\,P^i_{\mu\nu\rho\sigma}}{4\,\pi^2\left(x-y\right)^2}
	+ \mathcal O(|x-y|)
	\ ,
	\label{Gret}
\end{align}
where $\ell^2\equiv \ell^2(x,y)=\eta_{\mu\nu}\,\dd x^\mu\, \dd x^\nu$ is the background proper distance.
The contraction $P_{\mu\rho\ \ \nu}^{i\ \ \rho}\,\dd x^\mu\, \dd x^\nu$ will always result
in a factor of $\ell^2$ in the numerator that can potentially be canceled by a divergence
$\ell^{-2}$ of the propagator, leaving a non-zero minimum length behind.
Eq.~\eqref{eq:ds} finally becomes
\begin{equation}
	\lim_{x\to y}
	\ell_{\text{in-}\tau}^2
	=
	\begin{cases}
		0
		&
		\quad
		\tau = \text{in}
		\\
		0
		&
		\quad
		\tau = {\rm out \quad and \quad} \sum_i(\alpha_i+4\beta_i+\gamma_i) \leq 0
		\\
		\frac{2}{\pi}\,\lp^2\,
		\sum_i
		\left(\alpha_i + 4\,\beta_i + \gamma_i
		\right)
		\sim
		\lp^2
		&
		\quad
		\tau = {\rm out \quad and \quad} \sum_i(\alpha_i+4\,\beta_i+\gamma_i) > 0
		\ .
	\end{cases}
	\label{eq:zerolength}
\end{equation}
Therefore, quantizing gravity as a quantum field theory shows no sign of a minimum geometrical length,
but a minimum Planckian length scale is possible whenever $\sum_i(\alpha_i+4\,\beta_i+\gamma_i) > 0$.
We stress that the above result concerns only general properties of propagators and no mention
is made to specific models.
The information about particular theories is contained solely in the parameters $\alpha_i$, $\beta_i$ and
$\gamma_i$.
\par
In general relativity, for example, the massless spin-2 field (graviton)
is the only degree of freedom,
\beq
\hbar^{-1}
\Delta_{\mu\nu\rho\sigma}
=
\frac{\eta_{\rho\mu}\,\eta_{\sigma\nu}+\eta_{\sigma\mu}\,\eta_{\rho\nu}-\eta_{\mu\nu}\,\eta_{\rho\sigma}}
{q^2}
\ .
\eeq
In this case, $\sum_i(\alpha_i^{\rm GR}+4\,\beta_i^{\rm GR}+\gamma_i^{\rm GR}) = 4$ and a minimum length scale exists
which is given by
\beq
\ell_{\text{in-out}}^{\rm GR}(x,x)
=
\sqrt{\frac{8}{\pi}} \, \lp
\ .
\label{Lscale}
\eeq
Another interesting example is Stelle's theory~\cite{Stelle:1977ry,Stelle:1976gc},
whose spectrum contains additional degrees of freedom which are needed to prove the renormalisability.
In this case, the propagator reads
\beq
\hbar^{-1}
\Delta_{\mu\nu\rho\sigma}
=
\frac{2\,P^{(2)}_{\mu\nu\rho\sigma}- P^{(0)}_{\mu\nu\rho\sigma}}{q^2}
-\frac{2\, P^{(2)}_{\mu\nu\rho\sigma}}{q^2-m_2^2}
+ \frac{ P^{(0)}_{\mu\nu\rho\sigma}}{q^2-m_0^2}
\ ,
\label{eq:hdprop}
\eeq
where $P^{(s)}_{\mu\nu\rho\sigma}$ are spin-projection operators,
and one can see the additional massive degrees of freedom,
namely a scalar excitation of mass $\hbar\,m_0$ and a spin-2 particle
of mass $\hbar\,m_2$.
Surprisingly, due to the accidental cancelations of the parameters
\begin{equation}
	\sum_i \left(\alpha_i^{\rm St} + 4\,\beta_i^{\rm St} + \gamma_i^{\rm St} \right) = 0
	\ ,
\end{equation}
a minimum length scale does not exist, since Eq.~\eqref{eq:zerolength} yields
\beq
\ell_{\text{in-out}}^{\rm St}(x,x)
=
0
\ .
\label{eq:cancelations}
\eeq
\par
One should note that, unlike renormalisable theories that possess no natural scale,
non-renor\-malisable theories always come accompanied by an intrinsic scale $L_C$
which is then used to define the effective field theory.~\footnote{One should further
notice the difference between the role played by the intrinsic cutoff scale in non-renormalisable
theories and the scales obtained by dimensional transmutation, such as $\Lambda_{\rm QCD} = 220\,$MeV
in QCD.
The former represents the scale where the effective field theory breaks down
(and a more fundamental theory must be found).
Indeed, effective theories are defined as a series in inverse powers of the cutoff,
thus they are only meaningful at scales below it.
The latter, on the other hand, separates different regimes of the same theory,
which does not break down on either regime except, possibly, at some Landau pole.
We remark that the Landau pole of QED is way beyond the Planck scale,
thus it is automatically cured by the existence of a minimum length scale.}
For lengths $L \sim L_C$, the effective field theory breaks down, thus $L_C$ ``blinds''
all phenomena below it.
Therefore, within a non-renormalisable theory, $L_C$ serves as a kind of a minimum
length scale.
In general relativity, which is non-renormalisable, we indeed find the minimum scale $L_C\sim \ell_p$,
which turns out to be the same scale used to perform the effective field theory expansion.
On the other hand, Stelle's theory is renormalisable and should not (need or) provide any intrinsic scale.
The theory indeed knows nothing about the scale where it should fail and, if it were not for the ghost particle,
it could be extended to arbitrary scales.
Correspondingly, our calculation shows that Stelle's theory possesses no minimum scale.
This suggests an interesting interplay, perhaps a correspondence, between the renormalisability
of a theory (of gravity) and the non-existence of a minimum length scale.
\par
In a different perspective, like the one assumed in the asymptotic safety scenario~\cite{AS}
and classicalisation~\cite{classicalisation}, one might even argue that the length scale $L_C$ does not require
a new effective theory but that the (effective) theory is self-complete and simply rearranges
its degrees of freedom so that no new physics appears in the UV below $L_C$.
In particular, the minimum length scale $\ell_p$ in general relativity can be used to treat
its corresponding quantum field theory as fundamental, rather than effective, 
because $\ell_p$ plays the role of a natural regulator that removes all UV
divergences~\cite{pauli,Padmanabhan:1985jdl,Casadio:2008ug}.
Differently from a hard cutoff imposed by hand, which should be removed after renormalisation,
$\ell_p$ remains finite.
From this viewpoint, general relativity does not fail at the Planck scale, but rather physics beyond $\ell_p$
becomes (operationally) meaningless.
\section{Minimum length scale and GUP}
\label{S:gup}
\setcounter{equation}{0}
The famous {\em gedanken\/} experiment of the Heisenberg microscope~\cite{heisenberg}
shows that scattering processes are generically involved in quantum mechanical measurements.
Heisenberg's original idea was to measure position and momentum of a static particle,
say an electron, by using a photon as a probe.
The photon scatters off the electron, and by measuring the properties of the photon after the scattering,
one would like to know the exact position $x_{\rm e}$ and momentum $p_{\rm e}$
of the electron at the instant of the scattering.
However, since the photon has a wavelength $\lambda$, from the principles of wave optics follows
that the uncertainty in the position of the electron is (at least) $\Delta x_{\rm e} \simeq \lambda$.
Moreover, the photon carries a momentum $p=h/\lambda$, which, during the scattering, is partially
transferred to the electron in an unknown magnitude and direction.
This implies that, just after the scattering, the uncertainty in the electron momentum amounts to
(at most) $\Delta p_{\rm e} \simeq p = h/\lambda$. 
Therefore, Heisenberg concluded that
\be
\Delta x_{\rm e}\, \Delta p_{\rm e}
\simeq
\lambda \cdot \frac{h}{\lambda}
\simeq
h
\ .
\ee         
Successively, Schr\"odiger and Robinson formulated the uncertainty principle for canonically
conjugated variables, such as the position $x$ and momentum $p$ of a particle, in the form
\be
\Delta x \Delta p \geq \frac{\hbar}{2}
\ ,
\label{hep}
\ee
which is the one commonly accepted today. 
\par 
Heisenberg's heuristic approach paved the way to the formulation of
GUPs~\cite{Hossenfelder:2012jw,GUP1,GUP2,GUP3,Casadio:2020rsj}, which originate
from taking into account the gravitational effects in the photon-particle interaction.
For example, (microscopic) black hole formation in the measuring process (or the photon-electron
gravitational attraction) implies the existence of a minimum testable length
below which position measurements become meaningless.
Such GUPs can be mathematically encoded in modified quantum mechanical commutators,
and there is the tendency to extend such modification to quantum field theory commutators.
However, in the previous section we showed that a minimum length scale can be obtained without
modifying the quantum field theory dynamics.
This point of view implies that any GUP should emerge effectively in quantum mechanics as
the non-relativistic sector of quantum field theory of gravity without modifying the field propagators.
\par
The simplest form of GUP is given by
\be
\Delta x\,\Delta p
\ge
\frac{\hbar}{2}
\left(1+\frac{\delta_0}{\mpl^2}\,\Delta p^2\right)
\ ,
\label{dxdp}
\ee
where $\Delta O^2\equiv\expec{\hat O^2}-\expec{\hat O}^2$ for any quantum observables $\hat O$
and $\delta_0$ is a dimensionless deforming parameter expected to emerge
from candidate theories of quantum gravity.
Uncertainty relations can be associated with (fundamental) commutators
by means of the general inequality
\be
\Delta A \,\Delta B
\geq
\frac{1}{2}\left|\expec{[\hat A, \hat B]}\right|
\ .
\label{gen}
\ee
For instance, one can derive Eq.~\eqref{dxdp} from the commutator~\footnote{We stress
that the derivation of Eq.~\eqref{dxdp} only makes use of the algebraic structure of the commutator~\eqref{gup}
through the general inequality~\eqref{gen} and no specific representation of the physical operators $\hat{x}$ and
$\hat{p}$ (in whatsoever form) is needed.}
\be
\left[\hat x, \hat p\right]
=
i\,\hbar
\left(1 + \frac{\delta_0}{\mpl^2}\, \hat p ^2\right)
\ ,
\label{gup}
\ee
for which Eq.~\eqref{gen} yields
\be
\Delta x\,\Delta p
\ge
\frac{\hbar}{2}
\left[
1+\frac{\delta_0}{\mpl^2}
\left(\Delta p^2+\expec{\hat p}^2
\right)
\right]
\ .
\label{sb}
\ee
This immediately implies that the GUP~\eqref{dxdp} holds for any quantum
state, since $\expec{\hat p}^2 \geq 0$.
In particular, in the centre-of-mass frame of a scattering process, one can just consider the so-called
mirror-symmetric states satisfying $\expec{\hat p} = 0$,
and the inequality~\eqref{sb} coincides with the GUP~\eqref{dxdp}.
\par
Eq.~\eqref{dxdp} implies the existence of a minimum (effective) length
\be
\ell=\lp\,\sqrt{\delta_0}
\ .
\ee
Therefore, by comparing with Eq.~\eqref{eq:zerolength} from the previous section, we obtain
\be
\delta_0
=
\frac{2}{\pi}\,
\sum_i
\left(\alpha_i + 4\,\beta_i + \gamma_i
\right)
\ ,
\label{d0}
\ee
namely we arrive at an exact expression for the deformation parameter of the GUP
which should hold for a general class of gravity theories. 
\par
We can estimate $\delta_0$ for various models.
For general relativity, for example, one finds
\begin{equation}
	\delta_0^{\rm GR} = \frac{8}{\pi}
	\ ,
	\label{deltaGR}
\end{equation}
whereas for Stelle's theory we have
\begin{equation}
	\delta_0^{\rm St} = 0
	\ .
\end{equation}
We recall that, exact values of the deformation parameter $\delta_0$ had already been obtained in the past
from different approaches.
For example, it was found that $\delta_0=82\,\pi/5$ for general relativity in Ref.~\cite{SLV}
and $\delta_0=8\,\pi^2/9$ for models involving a maximal acceleration in Ref.~\cite{GPLPetr}.
All available results hence agree in order of magnitude.
Experimental upper bounds on $\delta_0$ exist~\cite{tests} (see also references therein), but they are
typically too weak ($\delta_0 \lesssim 10^{36}$) to provide any useful information about the gravitational propagator.
On the other hand, the theoretical value given by Eq.~\eqref{d0} can be viewed as a (general)
lower bound.
\section{Renormalisability and the sign of $\delta_0$}
\label{signdelta}
\setcounter{equation}{0}
From the discussion at the end of Section~\ref{S:minlength}, one could conjecture a correspondence
between the renormalisability of a theory and the absence of a minimum length scale in such a theory.
In terms of $\delta_0$, according to Eqs.~\eqref{eq:zerolength} and \eqref{d0}, such correspondence
would require
\begin{equation}
\delta_0 \leq 0
\end{equation}
for a renormalisable theory of gravity.
Interestingly, there are studies which consider this possibility for the GUP~\cite{Casadio:2020rsj,negGUP2,negGUP1,negGUP3}.
\par
In particular, the fundamental commutator was computed on a discrete lattice,
to our knowledge, for the first time in Ref.~\cite{negGUP2}.
This construct can in principle be viewed as a crystal-like model of our Universe, the so called ``world crystal'',
when the lattice spacing $\varepsilon$ is of the order of the Planck length.
The commutator then reads
\be
[\hat{x},\hat{p}]
=
i\,\hbar\,
\cos\left(\frac{\varepsilon}{\hbar}\,\hat{p}\right)
\ .
\label{latcomm}
\ee
At low energies, or momenta $|p|\ll \hbar/\varepsilon$, Eq.~\eqref{latcomm} implies
\be
\Delta x\, \Delta p
\geq
\frac{\hbar}{2}
\left(
1 - \frac{\varepsilon^2}{2\,\hbar^2}\,\Delta p^2
\right)
\ ,
\ee
where a negative $\delta_0 \equiv -\varepsilon^2\,\mpl^2/2\,\hbar^2\sim -(\varepsilon/\lp)^2$ can be clearly identified.
For large momenta approaching $p\simeq\pi\,\hbar/2\,\varepsilon\sim \mpl\,(\lp/\varepsilon)$, Eq.~\eqref{latcomm}
instead yields
\be
\Delta x\, \Delta p
\geq
\frac{\hbar}{2}
\left(\frac{\pi}{2} - \frac{\varepsilon}{\hbar}\,\expec{\hat p}
\right)
\simeq 0
\ .
\ee 
This result shows that no strictly positive lower bound for the uncertainty of two conjugate observables appears
when the energy reaches Planckian scale (for $\varepsilon\sim\lp$).~\footnote{One may say that the world-crystal
Universe appears to become ``classical'' and ``deterministic'' in this Planck regime.}
\par
The above example allows us to provide an alternative formulation of
our conjectured correspondence between renormalisability and the absence of a minimum length scale.
In fact, the theory in Ref.~\cite{negGUP2} is defined on a lattice so as to be finite for any
values of $\varepsilon\ge 0$.
Hence, it is renormalisable (in the limit $\varepsilon\to 0$) {\em by construction\/}
and displays a parameter $\delta_0\sim -\varepsilon\le 0$ corresponding to the absence of a GUP 
minimum length scale for any values of $\varepsilon\ge 0$.
Of course, \textit{any} quantum field theory should be UV finite when regularised on a lattice with a finite
step $\varepsilon>0$. 
However, this does not imply that the theory is renormalisable in the (continuum) limit $\varepsilon\to 0$.
According to our conjecture, this should occur if the GUP parameter $\delta_0$ computed in the regularised
theory does not become positive for the (lattice) regulator $\varepsilon\to 0$.
\par
Note that our conjecture makes non-renormalisable theories as good as renormalisable ones.
Indeed, the presence of a minimum length scale makes the former self-complete, which does not require
new physics beyond the Planck scale.
From this perspective, non-renormalisable theories are, in particular, finite and
predictive~\cite{pauli,Padmanabhan:1985jdl,Casadio:2008ug}.
On the other hand, we also stress that there is nothing worrisome about a non-positive GUP parameter,
thus renormalisable theories remain good candidates for describing fundamental physics.
\par
From a physical point of view, the sign of $\delta_0$ can give rise to a rich and varied phenomenology.
A positive $\delta_0$ is consistent with results obtained from {\em gedanken\/} experiments in
high-energy string scatterings, which also suggest the existence of an effective minimum length.
Furthermore, a GUP with a positive deforming parameter can play an important role in the Hawking
radiation~\cite{Hawking:1975vcx}.
From Heisenberg's uncertainty relation~\eqref{hep} [corresponding to~\eqref{dxdp} with $\delta_0=0$],
one can show that the temperature of a spherically symmetric black hole blows up as its mass decreases
during the evaporation~\cite{Fabio1995}.
This is in agreement with Hawking's original analysis which predicts that black holes should evaporate completely 
in a finite amount of time by reaching zero mass at infinite temperature.
Instead, from a GUP with positive $\delta_0$, one finds that the evaporation process would end in a finite time
with a remnant of finite mass and finite final
temperature~\cite{Fabio1995,ACS,Chen:2004ft}.~\footnote{A final positive temperature may sound
puzzling.
However, it has been shown within the GUP scenario that the specific heat of the remnant drops to zero when
its mass reaches the minimum value (see, e.g.~Refs.~\cite{ACS,Cavaglia,Scardigli}).
This result signals that any thermodynamic interaction of the black hole with the environment stops when the minimum
mass is reached.
It should then be possible to infer that the remnant temperature drops swiftly to zero in the very
last instants of the black hole life.
On this specific aspect, we can add that the evaporation models studied for example in Refs.~\cite{Nicolini,BR}
are more refined than those from the GUP (with $\delta_0>0$), as they analytically predict not only a final finite
mass of the remnant, but also a final temperature equal to zero. Lastly, we should notice that a vanishing final
temperature agrees with the requirement of total energy conservation realised by employing a microcanonical
description of the Hawking radiation~\cite{Hawking:1976de} and black hole microstates~\cite{Casadio:1997yv}.}
This result could have significant physical implications as, for example, black hole remnants are
considered for potential candidates of dark matter (see, e.g.~Ref.~\cite{Chen:2004ft}).
The existence of such remnants would also avoid issues like the information loss problem~\cite{Mathur:2009hf}, 
but would raise the question of their detectability and how to avoid their excessive
production in the early universe~\cite{Giddings:2008gr}.
On the contrary, an interesting implication of a GUP with $\delta_0<0$ is a finite final Hawking temperature and
a zero mass remnant at the end of the evaporation process (see Refs.~\cite{negGUP2,Ong2}~\footnote{There,
it was shown that the evaporation ends with final zero mass, when the temperature approaches a maximum
positive value, at which point the specific heat also drops to zero [see Eqs.~(72)-(73) in Ref.~\cite{negGUP2}].
Besides, a vanishing total energy is a feature of the classical model of point-like particles introduced in
Ref.~\cite{Arnowitt:1960zz} (for its quantum version, see Refs.~\cite{Casadio:2009jc}).}),
which would avoid
at once difficulties like the entropy/information problem, the remnant detectability issue, or their excessive production.
Further evidence in favor of a negative deforming parameter $\delta_0$ is the fact that this choice would resolve
the puzzle of white dwarfs by avoiding white dwarfs of arbitrarily large mass~\cite{negGUP1}. 
Finally, it was shown in Ref.~\cite{negGUP3} that the equivalence between the frameworks of
\textit{corpuscolar gravity\/} and GUPs also suggests a deforming parameter $\delta_0 < 0$,
once the usual energy conservation is imposed.
\section{Conclusions}
\label{conc}
\setcounter{equation}{0}
In this paper, we first reviewed the idea of a minimum geometrical length in quantum gravity defined by
in-in amplitudes obtained via the Schwinger-Keldysh formalism~\cite{Casadio:2020hzs}.
The in-in quantum proper distance can be interpreted as a truly geometrical length that
happens to be real at all loop orders and satisfies a causal equation of motion.
At coinciding points, the in-in proper length goes to zero at second order for any metric theory
of gravity, a result that extends to all orders in perturbation theory as long as
non-gravitational interactions can be neglected.
\par
On the contrary, a minimum length scale arises from the in-out amplitudes used to derive the standard
Feynman rules and propagators, and its value depends on parameters determined by the particular quantum
field theory of gravity considered.
We stress that this conclusion is a general, model-independent property.
For general relativity, viewed as a non-renormalisable field theory, our analysis implies the existence of a minimum
length scale of the order of the Planck length, as expected.
In Stelle's theory of gravity, which is renormalisable, we instead showed the absence of a minimum length scale.
An interesting interplay between the non-renormalisability of a (gravitational) theory and the existence of a minimum
length scale thus seems to emerge, the latter being the embedded (self-)cure for the former~\cite{pauli}.
\par
Minimum length scales are also widely described through the so called GUPs.
The minimum length scales derived from GUPs and those derived from in-out amplitudes
should be the same.
This suggests a deep connection between the parameters of quantum theories of gravity and the deforming parameter
of GUPs emerging in the non-relativistic limit of quantum field theory.
Using this identification, we find a GUP deforming parameter $\delta_0>0$ and of order unity (in Planck length)
for general relativity, consistently with previous evaluations~\cite{SLV,GPLPetr} and with some models of
string theory~\cite{GUP2}.
The  GUP deforming parameter for Stelle's theory instead vanishes.
As another example in support of the conjecture that a minimum length scale is induced by the non-renormalisability
of a field theory, we discussed the GUP formulated on a lattice, where a negative deformation parameter emerges
at low energy and vanishes at high energy~\cite{negGUP2}.
The fact that $\delta_0\le 0$ and no minimum length scale exists agrees with the expectation that any field theory
(of gravity) must be UV finite, hence renormalisable, if defined on a lattice. 
On the other hand, when the lattice acts as a regulator for UV divergences, we expect that the theory is 
renormalisable if the GUP parameter is not positive in the continuum limit.   
\par
The connection between quantum field theories and the deforming GUP parameter $\delta_0$ could be used,
in principle, to rule out some theories by measuring the value of the deforming parameter from experiments.
Unfortunately, current experimental upper bounds on $\delta_0$ are still way too weak to provide any useful information
on the underlying gravitational theory~\cite{tests}.
\section*{Acknowledgments}
R.C. and W.F.~are partially supported by the INFN grant FLAG.
The activity of R.C.~has also been carried out in the framework
of the National Group of Mathematical Physics (GNFM, INdAM).
W.F.~acknowledges the financial support provided by the scholarship
granted by the Chinese Scholarship Council (CSC).
%
%
%
%
%
%

%

\begin{thebibliography}{99}
%
\bibitem{pauli}
W.~Pauli, Letter of Heisenberg to Peierls (1930), in
{\em Scientific Correspondence}, editor K.~von
Meyenn (Springer-Verlag, 1985) p.~15, Vol~II;
O.~Klein,
Helv. Phys. Acta. Suppl. {\bf 4} (1956) 58.
%
\bibitem{Hossenfelder:2012jw}
S.~Hossenfelder,
Living Rev. Rel. \textbf{16} (2013) 2
[arXiv:1203.6191 [gr-qc]].
%
\bibitem{Padmanabhan:1985jq}
T.~Padmanabhan,
Gen. Rel. Grav. \textbf{17} (1985) 215.
%
\bibitem{Keldysh:1964ud}
L.~V.~Keldysh,
Zh. Eksp. Teor. Fiz. \textbf{47} (1964) 1515.
%
\bibitem{Jordan:1986ug}
R.~D.~Jordan,
Phys. Rev. D \textbf{33} (1986) 444.
%
\bibitem{Casadio:2020hzs}
R.~Casadio and I.~Kuntz,
Eur. Phys. J. C \textbf{80} (2020) 958
[arXiv:2006.08450 [hep-th]].
%
\bibitem{GUP1}
M.~P.~Bronstein, 
Phys. Zeitschr. der Sowjetunion {\bf 9} (1936) 140;
H.~S.~Snyder,
Phys. Rev. {\bf 71} (1947) 38;
C.~N.~Yang,
Phys. Rev. {\bf 72} (1947) 874;
C.~A.~Mead,
Phys. Rev. {\bf 135} B (1964) 846;
F.~Karolyhazy,
Nuovo Cim. A {\bf 42} (1966) 390.
%
\bibitem{GUP2}
D.~Amati, M.~Ciafaloni, G.~Veneziano,
Phys. Lett. B {\bf 197} (1987) 81;
D.~J.~Gross, P.~F.~Mende,
Phys. Lett. B {\bf 197} (1987) 129;
D.~Amati, M.~Ciafaloni, G.~Veneziano,
Phys. Lett. B {\bf 216} (1989) 41;
K.~Konishi, G.~Paffuti, P.~Provero,
Phys. Lett. B {\bf 234}(1990) 276.
%
\bibitem{GUP3}
M.~Maggiore,
Phys.\ Lett.\ B {\bf 304} (1993) 65
[hep-th/9301067];
%
A.~Kempf, G.~Mangano and R.~B.~Mann,
Phys.\ Rev.\ D {\bf 52} (1995) 1108
[hep-th/9412167];
F.~Scardigli,
Phys.\ Lett.\ B {\bf 452} (1999) 39
[hep-th/9904025];
%
R.~J.~Adler and D.~I.~Santiago,
Mod.\ Phys.\ Lett.\ A {\bf 14} (1999) 1371
[gr-qc/9904026];
%
S.~Capozziello, G.~Lambiase and G.~Scarpetta,
Int.\ J.\ Theor.\ Phys.\  {\bf 39} (2000) 15
[gr-qc/9910017];
%
P.~Bosso,
Class. Quant. Grav. \textbf{38} (2021) 075021
[arXiv:2005.12258 [gr-qc]].
%
\bibitem{Casadio:2020rsj}
R.~Casadio and F.~Scardigli,
Phys. Lett. B \textbf{807} (2020) 135558
[arXiv:2004.04076 [gr-qc]].
%
\bibitem{Nink:2014yya}
A.~Nink,
Phys. Rev. D \textbf{91} (2015) 044030
[arXiv:1410.7816 [hep-th]].
%
\bibitem{Demmel:2015zfa}
M.~Demmel and A.~Nink,
Phys. Rev. D \textbf{92} (2015) 104013
[arXiv:1506.03809 [gr-qc]].
%
\bibitem{Nink:2015lmq}
A.~Nink and M.~Reuter,
JHEP \textbf{02} (2016) 167
[arXiv:1512.06805 [hep-th]].
%
\bibitem{Stelle:1977ry}
K.~S.~Stelle,
Gen. Rel. Grav. \textbf{9} (1978) 353.
%
\bibitem{Stelle:1976gc}
K.~S.~Stelle,
Phys. Rev. D \textbf{16} (1977) 953.
%
\bibitem{AS}
A.~Bonanno, A.~Eichhorn, H.~Gies, J.~M.~Pawlowski, R.~Percacci, M.~Reuter, F.~Saueressig and G.~P.~Vacca,
Front. in Phys. \textbf{8} (2020) 269
[arXiv:2004.06810 [gr-qc]];
A.~Eichhorn,
Front. Astron. Space Sci. \textbf{5} (2019) 47
[arXiv:1810.07615 [hep-th]].
%
\bibitem{classicalisation}
G.~Dvali, G.~F.~Giudice, C.~Gomez and A.~Kehagias,
JHEP \textbf{08} (2011) 108
[arXiv:1010.1415 [hep-ph]];
R.~Percacci and L.~Rachwal,
Phys. Lett. B \textbf{711} (2012) 184
[arXiv:1202.1101 [hep-th]].
%
\bibitem{Padmanabhan:1985jdl}
T.~Padmanabhan,
Annals Phys. \textbf{165} (1985) 38.
%
\bibitem{Casadio:2008ug}
R.~Casadio,
Int. J. Mod. Phys. A \textbf{27} (2012) 1250186
[arXiv:0806.0501 [hep-th]].
%
\bibitem{heisenberg}
W.~Heisenberg,
``The Physical Principles of the Quantum Theory,''
(University of Chicago Press, Chicago, 1930).
%
\bibitem{SLV}
F.~Scardigli, G.~Lambiase and E.~Vagenas,
Phys. Lett. B \textbf{767} (2017) 242
[arXiv:1611.01469 [hep-th]].
%
\bibitem{GPLPetr}
G.~G.~Luciano and L.~Petruzziello,
Eur. Phys. J. C \textbf{79} (2019) 283
[arXiv:1902.07059 [hep-th]].
%
\bibitem{tests}
%
F.~Scardigli and R.~Casadio,
Eur. Phys. J. C \textbf{75} (2015) 425
[arXiv:1407.0113 [hep-th]];
%
F.~Scardigli,
J. Phys. Conf. Ser. \textbf{1275} (2019) 012004
[arXiv:1905.00287 [hep-th]];
%
M.~Hemeda, H.~Alshal, A.~F.~Ali and E.~C.~Vagenas,
``Gravitational Observations and LQGUP,''
[arXiv:2208.04686 [gr-qc]];
%
S.~Aghababaei, H.~Moradpour and E.~C.~Vagenas,
Eur. Phys. J. Plus \textbf{136} (2021) 997
[arXiv:2109.14826 [gr-qc]].
%
\bibitem{negGUP2}
P.~Jizba, H.~Kleinert and F.~Scardigli,
Phys. Rev. D \textbf{81} (2010) 084030
[arXiv:0912.2253 [hep-th]].
%
\bibitem{negGUP1}
Y.~C.~Ong,
JCAP \textbf{09} (2018) 015
[arXiv:1804.05176 [gr-qc]].
%
\bibitem{negGUP3}
L.~Buoninfante, G.~G.~Luciano and L.~Petruzziello,
Eur. Phys. J. C \textbf{79} (2019) 663
[arXiv:1903.01382 [gr-qc]].
%
%
\bibitem{Hawking:1975vcx}
S.~W.~Hawking,
Commun. Math. Phys. \textbf{43} (1975) 199
[erratum: Commun. Math. Phys. \textbf{46} (1976) 206].
%
\bibitem{Fabio1995}
F.~Scardigli,
Nuovo Cim. B \textbf{110} (1995) 1029.
%
\bibitem{ACS}
R.~J.~Adler, P.~Chen and D.~I.~Santiago,
Gen. Rel. Grav. \textbf{33} (2001) 2101
[arXiv:gr-qc/0106080 [gr-qc]].
%
\bibitem{Chen:2004ft}
P.~Chen,
New Astron. Rev. \textbf{49} (2005) 233
[arXiv:astro-ph/0406514 [astro-ph]].
%
\bibitem{Cavaglia}
M.~Cavaglia and S.~Das,
Class. Quant. Grav. \textbf{21} (2004) 4511
[arXiv:hep-th/0404050 [hep-th]].
%
%
\bibitem{Scardigli}
F.~Scardigli,
Symmetry \textbf{12} (2020) 1519
[arXiv:0809.1832 [hep-th]].
%
\bibitem{Nicolini}
P.~Nicolini,
Int. J. Mod. Phys. A \textbf{24} (2009) 1229
[arXiv:0807.1939 [hep-th]].
%
\bibitem{BR}
A.~Bonanno and M.~Reuter,
Phys. Rev. D \textbf{62} (2000) 043008
[arXiv:hep-th/0002196 [hep-th]].
%
\bibitem{Hawking:1976de}
S.~W.~Hawking,
Phys. Rev. D \textbf{13} (1976) 191.
%
\bibitem{Casadio:1997yv}
R.~Casadio and B.~Harms,
Phys. Rev. D \textbf{58} (1998) 044014
[arXiv:gr-qc/9712017 [gr-qc]].
%
\bibitem{Mathur:2009hf}
S.~D.~Mathur,
Class. Quant. Grav. \textbf{26} (2009) 224001
[arXiv:0909.1038 [hep-th]];
J.~D.~Bekenstein,
Phys. Rev. D \textbf{49} (1994) 1912
[arXiv:gr-qc/9307035 [gr-qc]].
%
\bibitem{Giddings:2008gr}
S.~B.~Giddings and M.~L.~Mangano,
Phys. Rev. D \textbf{78} (2008) 035009
[arXiv:0806.3381 [hep-ph]];
K.~Nozari,
Astropart. Phys. \textbf{27} (2007) 169
[arXiv:hep-th/0701274 [hep-th]];
A.~Barrau, C.~Feron and J.~Grain,
Astrophys. J. \textbf{630} (2005) 1015
[arXiv:astro-ph/0505436 [astro-ph]].
%
%
\bibitem{Ong2}
Y.~C.~Ong,
JHEP \textbf{10} (2018), 195
[arXiv:1806.03691 [gr-qc]].
%
\bibitem{Arnowitt:1960zz}
R.~Arnowitt, S.~Deser and C.~W.~Misner,
Phys. Rev. Lett. \textbf{4} (1960) 375.
%
%
\bibitem{Casadio:2009jc}
R.~Casadio, R.~Garattini and F.~Scardigli,
Phys. Lett. B \textbf{679} (2009) 156
[arXiv:0904.3406 [gr-qc]];
R.~Casadio,
Int. J. Mod. Phys. A \textbf{28} (2013) 1350088
[arXiv:1303.1274 [gr-qc]].
%
%
\end{thebibliography}
\end{document}